# High Anisotropic Inserted Dendritic Growth During First Stage of Zn Monolayer Deposition on Ag(111) Substrate

**Hicham Maradj[a], Nabil Rochdi[b], Benedicte Ealet[a], Sebastien Vizzini[c], Jean-Paul Biberian[a], Bernard Aufray[a], Haik Jamgotchian[a,*]**

[a] Aix Marseille Univ, CNRS, CINAM, Marseille, France

[b] Cadi Ayyad University, Faculty of Sciences Semlalia, Department of physics, Prince Moulay Abdellah avenue, PO Box 2390, 40000 Marrakesh, Morocco

[c] Aix Marseille Univ, CNRS, IN2MP, Marseille, France

* Corresponding author: jamgotchian@cinam.univ-mrs.fr

**Abstract**

Sub-monolayer physical layer deposition of Zn on Ag(111) substrate at room temperature is investigated using Auger electron spectroscopy, low energy electron diffraction and scanning tunneling microscopy. We identified an original Zn highly-anisotropic structure in the shape of finger-like inserted dendrite in front of Zn monolayer growth. On the basis of STM observations, a mechanism of inserted dendritic growth and a mechanism of its transition to layered growth are proposed.

## 1. Introduction

The deposition of metallic thin films, especially metallic monolayers, on crystalline substrates is of great interest because of their relevant potential applications in the domain of catalysis, magnetism and nanoelectronics for instance [1–5]. In addition, ordered or self-assembled structures involved in nanoscaled materials are attractive regarding the important relationship between their structures and their specific physical and chemical properties. More particularly, the deposition of a metal up to one monolayer takes its interest in view of the diversity of possible resulting structures such as islands, fractals or surface dendrites formation [6,7]. The understanding of mechanisms governing their formation is particularly essential for the elaboration of materials with one or more types of integrated order such as ferroics (*i.e.* magnetic, ferroelectric, and ferroelastic materials) and multiferroics (exhibiting more than one ferroic behavior) [8]. Thus, the understanding of the epitaxial growth is of great significance.

The growth of metallic films with a coverage exceeding the monolayer occurs following Frank van der Merwe (layer by layer), Volmer-Weber (island) or Stranski-Krastanov (layer and island) growth modes [9–12]. Generally, the deposition of a metal A on a metallic substrate B is dependent on the relative segregation phenomenon of both elements in the AB alloy, in particular when the substrate temperature allows atomic exchanges close to the surface. More precisely, the surface energy difference, the atomic radius difference and the order or phase separation tendency are the main driving forces to be considered in order to predict the element which tends to segregate at the surface [13,14]. In most cases, it is the coupling between these three competing effects that governs the behavior of the system. Using this approach, numerous experimental results on bimetallic systems have been interpreted giving rise to relevant predictions on the evolution of these systems [15].

From a structural point of view, the deposition of one monolayer of an element A with a lower surface energy, on an element B with a higher surface energy, typically results in an atomic arrangement which is strongly dependent on the difference between the atomic sizes of the two elements. In other words, when the atomic radius of the adsorbate (*i.e.* the deposited element) is larger than that of the substrate (which is commonly the case), the monolayer forms a Moiré pattern (MP) due to the Vernier effect, and which is directly defined by the difference between the adsorbate and the substrate atomic radii. However, when the adsorbate atoms have relatively the same radius as the substrate ones, a coherent epitaxy (so-called pseudomorphy) of the monolayer takes place (*i.e.* the monolayer of the deposited element presents the same atomic parameters than that of the substrate). As a matter of fact, a small difference between the adsorbate and the substrate atomic radii gives rise to pseudomorphy domains separated by periodic stacking defects (so-called quasi-pseudomorphy epitaxy). Since the interaction potential between two atoms is asymmetrical (*i.e.* the monolayer is energetically more extensible than compressible), a quasi-pseudomorphy structure is then expected when the adsorbate atoms are smaller than the substrate ones [16].

In principle, the deposition of one Zn monolayer on Ag(111) should correspond to this last structure due to the smaller size of Zn atoms when compared to Ag ones. Nevertheless, we have shown in a previous paper [17] that one Zn monolayer (ML) on Ag(111) substrate gives rise to a MP formation as expected when adsorbed atoms are larger than substrate ones; the deposition of one Zn ML corresponds to a coverage ($\theta$) of the Ag(111) surface of 1.18. Here, one Zn ML corresponds to the calculated quantity of Zn forming a trigonal patterned (111) monolayer.

Regarding the size effect, the Zn/Ag(111) is similar to the Co/Pt(111) system (*i.e.* when the atomic radius of the adsorbate is smaller than that of the substrate) for which the deposition at $\theta=1.07$ was shown to lead to the expected Co quasi-pseudomorphic layer, *i.e.* stable alternate trigonal domains of face-centered cubic (FCC) and hexagonal close-packed (HCP) structures separated by stacking faults [18].

Using Auger electron spectroscopy (AES), Maradj et al. have shown that the monitoring of the growth kinetics of Zn on Ag(111) (*i.e.* the variation of the Zn Auger peak intensity versus the deposition time) reveals an unexpected delay at the beginning of the growth [17]. A similar detection delay at the growth onset was observed by Kourouklis et al. on the same metallic system and in the same growth conditions, and was shown to increase with the sample temperature and with the evaporation rate at a given temperature [19]. Kourouklis et al. ascribed this behavior to the Volmer-Weber growth mode (*i.e.* the formation of 3D clusters on the surface) meanwhile it was interpreted by Maradj et al. as a dissolution process. Furthermore, using scanning tunneling microscopy (STM) and low energy electron diffraction (LEED), Maradj et al. observed that the delayed Zn detection is followed by the gradual formation of a Zn dense plane generating a MP, which proves that the growth occurs through the layer by layer growth mechanism. Therefore, the study of early stages of Zn growth on Ag(111) at room temperature (RT) is crucial to understand the mechanisms that take place.

In this paper, we report an original effect never seen before on the Zn/Ag(111) system prior to the formation of a Zn monolayer. It is about the formation of regular and highly-anisotropic nanometer-scale dendrites. These original dendrites are different from the dendrites of solidification by their size (2.5 nm instead of hundreds of microns) and from dendrites obtained by Diffusion Limited Aggregation by the fact that they are inserted in the surface of the substrate (and not onto the surface). In addition, this original dendritic growth presents an anisotropy regarding the orientation of the step where it occurs, which reflects another original aspect of the present work. The presented results are supported by STM measurements at RT of the Zn morphology on the Ag(111) surface obtained by the deposition of less than 1 ML. In particular, attention was paid to the mechanisms taking place during the very early stages of Zn growth. From a detailed STM analysis, we propose a growth mechanism of Zn atoms explaining the specificity of the Zn/Ag(111) system. In the following, we describe the experimental set-up and then we present and discuss the experimental results obtained by AES, LEED and STM.

## 2. Experimental set-up

The experiments are performed using the same ultra-high vacuum (UHV) equipment (with a base pressure better than $10^{-9}$ mbar) and following the same procedure described in [17]. The surface cleaning of the Ag(111) crystal is performed in a separated UHV preparation chamber by iterative sequences of ionic argon sputtering (at 550 eV for a duration of 1 h) followed by annealing in UHV conditions (at 800 K for a duration of 1 h). The cleanness of the Ag surface is checked by an Auger spectrometer (with a cylindrical mirror analyzer) and a LEED optics system in the main chamber, which is further equipped with a Zn evaporator used for the deposition. The latter consists of an alumina crucible (heated by a tungsten filament) which is degassed at low temperature during 30 min before Zn evaporation. The monitoring of the Zn growth kinetics is based on chemical and structural evolutions of the surface, and consists in systematic AES and LEED analyses after each Zn deposited amount. The calibration of deposited Zn amounts (*i.e.* the Zn evaporation rate) was performed on the basis of AES, LEED and STM measurements as previously reported [17]. The STM characterization is performed in a third chamber equipped with an Omicron STM-1 operating at room temperature with a tungsten tip. All the STM images are recorded in the filled states mode.

## 3. Results

### 3.1. AES-LEED study

Fig. 1 displays the intensities evolution of Zn (56 eV and 998 eV) and Ag (355 eV) Auger signals versus the Zn deposition time. One can clearly notice that the evolution of the Zn Auger peak intensity at low energy (solid squares) presents a delay at the beginning of the growth as it was previously observed [17]. After this delay, there is a quasi-linear increase of the Zn intensity up to a slope change (a break) at about 10 min of deposition. After this break, the Zn intensity increases slightly. Besides, the intensity of the Ag Auger peak (solid triangles) is continuously decreasing down to a value close to zero. From LEED and STM characterizations, we have shown that a dense compact Zn monolayer forming a Moiré pattern with the Ag(111) substrate is obtained after a duration of about 10 min of Zn deposition (pointed out by a black vertical line on Fig. 1). The Zn detection delay is about 2 min of Zn deposition time.

Indeed, the calibration of the Zn quantity has been done by the fact that after 10 min of deposition, a Zn(111) monolayer is formed on the substrate, which corresponds to $\theta=1.18$ (this value does not take into account the quantity of Zn dissolved into the substrate). Thus, the observed slope change at 10 min confirms the deposition of a Zn(111) monolayer in agreement with the calibration mentioned above. In this respect, it is worth to be noted that the Auger measurements revealed neither a prominent signature of oxygen originating from contaminants (*i.e.* from adventitious carbon and oxygen species inherently present even in UHV environments) nor oxygen-induced-shifts of the Zn peaks.

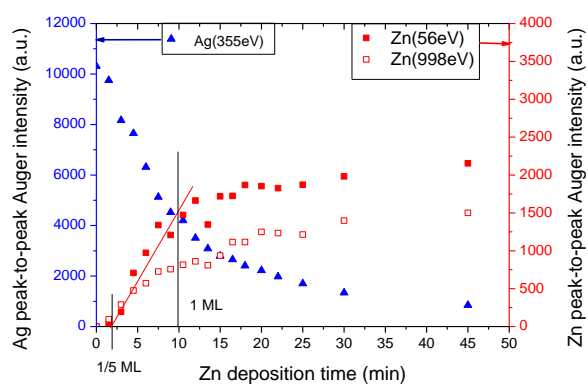

**Fig. 1.** Auger signal intensities of Zn (56 eV and 998 eV) and Ag (355 eV) as a function of the Zn deposition time at room temperature [17]. The vertical line at 10 min corresponds to the formation time of one Zn monolayer; the slope break of the Zn Auger signal proves the layer by layer growth. The vertical line at 2 min shows the beginning of the layered growth. The high energy peak of Zn (998 eV) shows exponential growth kinetics from the beginning of deposition.

In order to determine the physical origin of this detection delay (equivalent to a slowing down) observed at the beginning on the Auger growth kinetics (mainly observed on the Zn low energy Auger signal), we have investigated the surface morphology using STM during a first stage of the growth. In the following, we present the surface morphology observed after 3 min of Zn deposition, *i.e.* after the deposition of Zn corresponding to θ=0.35 (equivalent to 0.3 Zn ML). After this deposition time, no modification of LEED pattern was observed (*i.e.* the LEED pattern shows only the initial spots of the pristine Ag(111) surface).

**3.2. STM study**

After 3 min of Zn deposition, a typical morphology of the surface was observed by STM as reported on Fig. 2. On this STM image, one can distinguish three terraces with different Zn growth signatures. On the left part of the upper terrace, we observe two domains separated by a continuous and slightly bended line. On the lower side of this line, up to the step-edge, we observe a Moiré pattern (labelled MP on the STM image) corresponding to a complete ML of Zn as previously reported [17]. Above this line, we observe a terrace without MP, *i.e.* likely without Zn atoms onto or in the topmost surface layer (labelled Ag on the STM image). Nevertheless, a magnification of a part of this area (Fig. 2b) shows small unstructured dark areas with a very low corrugation (less than 25 pm), and which could be the signature of Zn atoms buried close to the Ag surface. This interpretation is relevant under the assumption that the upper surface consists of pure Ag without Zn elements.

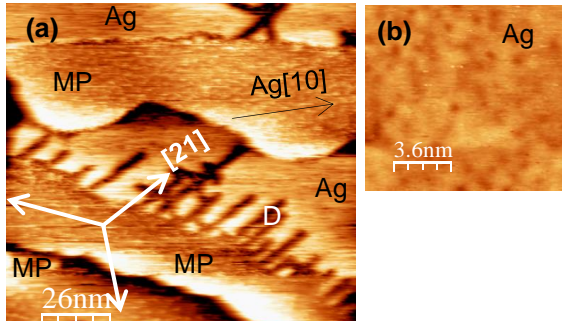

**Fig. 2.** (a) STM image recorded after the deposition of 0.3 Zn ML on the Ag(111) (U = 0.6V, I = 1.7 nA). The white arrows show the three dendritic growth directions. MP: Moiré pattern, D: Dendrite, Ag: Silver terrace; (b) Blow-up part of the STM image showing the MP-free Ag surface with corrugated dark areas.

On the middle terrace, we observe another growth front giving rise to a Moiré structure behind it. This growth front is headed by a set of dark linear fronts like fingers or surface dendrites (labelled D on the STM image). These dendrites are mainly oriented in the [21] direction of the Ag surface and they all have the same width (~2.5 nm). Similar but isolated dendrites are also observed in the above terrace, and are also located at the border between the MP domain and the Zn-free Ag area (pristine silver surface). The white arrows show the preferential directions of dendrites growth. Finally, one could remark that the lower terrace is completely covered by a Moiré pattern, and that the Moiré pattern of the middle terrace is less regular than in other terraces.

Fig. 3a shows another typical area of the sample which also exhibits different growth morphologies. On this STM image, the line labelled (1) is a 2D interface between a complete Zn ML (exhibiting the MP) and a pristine Ag(111) surface (*i.e.* Ag without Zn). The line labelled (2) is a step-edge formed by the ML of Zn and pure Ag(111). The depth profile shown in Fig. 3b corresponds to the green line on the STM image of Fig. 3a. The depth profile across the interface (1) shows a difference in height of about 35 pm. This value is similar to the size difference between Zn (266 pm) and Ag (289 pm) atoms (of about 25 pm) as shown on the hard spheres representation of this kind of interfaces (Fig. 3c). The corrugation profile across the dendrites (~35 pm) indicates that they are inserted (embedded) in the Ag(111) surface layer. Finally, the height difference of the step-edge between Ag and MP terraces (2) is about 0.2 nm, which is in agreement with the associated schematic diagram depicted in Fig. 3d.

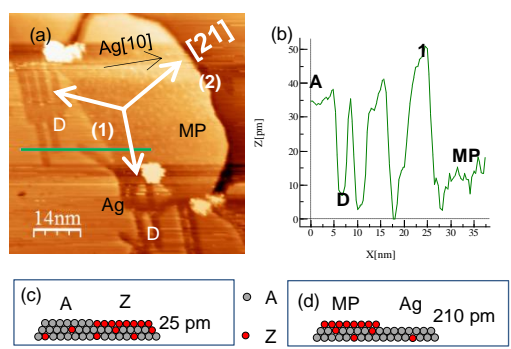

**Fig. 3.** (a) STM image showing dendrite (D) and Moiré pattern (MP) formation (U = 1.2 V, I = 0.8 nA). The white arrows show the three preferential dendritic growth directions; (b) Depth profile of the green line of (a); (c) Zn-Ag interface hard sphere presentation. The Ag layer is 25 pm higher than Zn one; (d) MP-Ag step-edge hard sphere presentation. The MP is 210 pm higher than the Ag terrace.

Therefore, we identified two different growth fronts of the MP: (i) a growth front in the dendrites side (*e.g.* interface (1) in Fig. 3a), and (ii) a growth front on a step-edge (*e.g.* interface (2) in Fig. 3a). On the STM image of Fig. 3, the white protrusions are likely impurities which have stopped or slowed down the growth of the Zn ML. This is in line with the fact that the Zn ML growth is mainly due to the advance of the step-edge type (2) rather than the advance of the interface type (1) (which corresponds likely to a slower growth). While the step-edge growth mechanism corresponding to the case (ii) is very common in the layered growth, the growth mechanism corresponding to the case (i) is unusual and deserves to be attentively examined. In this respect, we have noticed that the growth of dendrites strongly depends on the relative orientation of primary dendritic arms regarding the step-edge orientation. The STM images of Fig. 4 present the onset of dendritic growth with different orientations with respect to the step-edge. One can notice that dendrites formed by primary arms are perpendicular to the step-edge on Fig. 4a, while they have another orientation (different from the step-edge normal) on Fig. 4b. Between the dendritic structure and the step-edge, one can distinguish a quite narrow area which corresponds probably to the Moiré pattern.

On Fig. 4c, we can see another example where the angle between the step-edge and primary dendritic arms is about 30°. For this particular angle, there is a coexistence of two orientations of primary dendritic arms, with secondary arms growing in the back direction perpendicularly to the step-edge. The dendritic growth along these three directions brings about the formation of Ag islands (I) as a result of the intersection of dendrites arms. Besides, one could notice that, in spite of the equivalence between step-edges orientations (*e.g.* Ag [01] in Fig. 4a, and Ag [-1-1] in the Fig. 4c), dendritic growth orientations with respect to the step-edge are nonequivalent, indicating then the non-axisymmetric property of this dendritic growth. Note that the growth direction of all secondary arms is at 120° with respect to primary arms, which is not a trivial fact since it corresponds rather to a backward direction when compared to the advance direction of the Zn ML/Ag growth front (*e.g.* equivalent to the type of interface (1) in Fig. 3a). Besides, it is worth noting that these secondary arms do not grow starting from the MP front type (1) and they can cross other dendrites because of their orientation (we refer to such a crossing as secondary arms cross-back mechanism). Indeed, we can clearly identify such a mechanism through the association of an Ag island (I) with secondary dendritic arms (D) in Fig. 4c. Concomitantly, the step-edge growth (growth case (ii)) could occur in the opposite side.

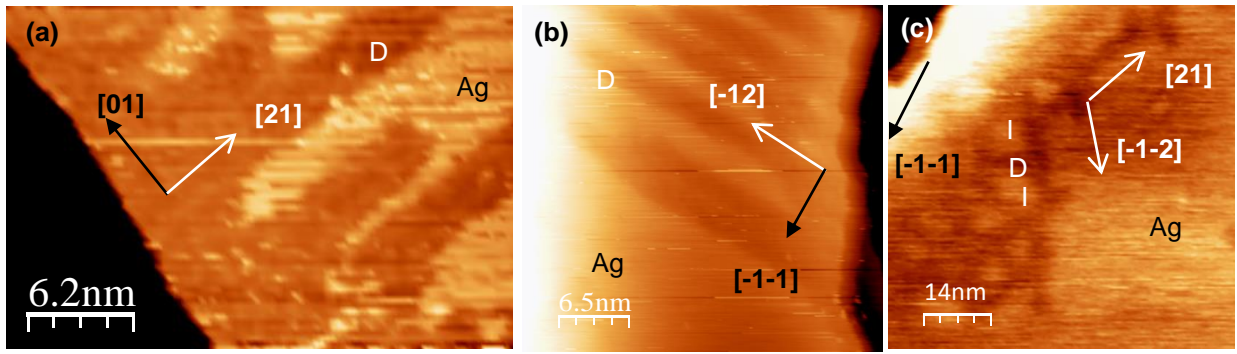

**Fig. 4.** Dendritic growth on different step-edge orientations. The white arrows show the primary dendritic arms growth directions. (a) The primary dendritic arms are perpendicular to the step-edge; (b) The primary dendritic arms forming an arbitrary angle with the step-edge; (c) The primary dendritic arms forming an angle of about 30° with the step-edge. D: Dendrite, Ag: Silver terrace, I: Ag island.

## 4. Discussion

### 4.1. Dendritic growth

Taking into account the three driving forces of the segregation phenomenon of Zn on Ag (the size effect, the surface energy effect and the alloy formation tendency), the growth mode of dendrites occurs probably according to two steps: The first step is a dissolution process of Zn atoms into the Ag surface selvedge (*i.e.* located close to the surface) and the second step consists in a segregation process of Zn atoms. The above-mentioned dissolution process corresponds to the delay (2 min) observed at the beginning of the Auger growth kinetics (Fig. 1), and is consistent with the quite large solubility of Zn in Ag of about 30% at RT [20]. Moreover, the small unstructured dark areas with a very low corrugation observed on pristine Ag terraces (Fig. 2b) could be the signature of Zn atoms buried close to the Ag surface. In the framework of this interpretation, Zn atoms do not form a solid solution with Ag but locally form quite disordered intermetallic compounds more probably.

Concerning the surprising dendrites (D) observed by STM in front of Zn ML forming a MP (Fig. 3), it should be emphasized that they all have the same width (2.5 nm) and appear only in a specific orientation. Furthermore, from a detailed analysis of different depth profiles, these dendrites were found to be embedded in the topmost Ag surface layer. The limited and constant width of these dendrites suggests a strain effect. The latter could arise from the Zn substitution (*i.e.* Zn atoms substitution for Ag ones) in the topmost surface layer of Ag such in a perfect epitaxy (*i.e.* there is an exchange between Ag and Zn atoms). In this case, the constant and limited width (2.5 nm) is the consequence of the size difference between Ag and Zn atoms inducing a stretch strain. On the contrarily, it is not possible to interpret the dendrites length on the basis of the strain induced by inserted Zn atoms. Indeed, the dendrites direction is the same as the Ag [21] direction, which rules out an equivalent strain relaxation via a Vernier effect since we had never observed a Moiré Pattern inside the dendrites (otherwise, the lateral strain would be also modified). Here we are dealing with an uncommon situation where relaxation of substitutional Zn atoms occurs only in one direction.

We think that this unusual behavior could be due to a specificity of the Zn bulk structure that we plotted in Fig. 5. Indeed, the atomic structure of bulk Zn is a hexagonal close packed structure with the lattice parameters a=0.266, b=0.266, and c=0.495 nm. Each Zn atom has 6 nearest neighboring atoms in the dense hexagonal plane (0001) and 6 others with the upper and lower dense planes. The dense plane of Zn monolayer has a trigonal isotropic structure with $d_1 = 0.266$ nm. However, the interatomic distance is different out of the dense plane ($d_2 = 0.291$ nm), which is close to the interatomic distance of Ag (0.289 nm). Thus, one can assume that a perfect hetero-epitaxy of Zn atoms takes place inside the dendrites with an asymmetric strain relaxation. One can also imagine a buckled plane of Zn(01-10) as drawn in purple on Fig. 5.

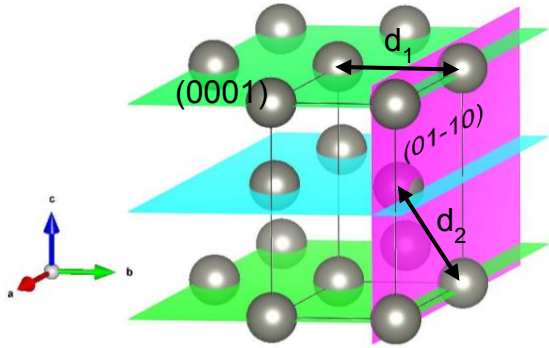

**Fig. 5.** Atomic presentation of the Zn unit cell in the hexagonal closest structure.

The specific observed directions of the dendrites growth (*i.e.* along the Ag[21] direction and the two other equivalent directions at +/- 120°) could be ascribed to an anisotropic growth due to the inequivalent directions [10] and [01], which is a specificity of the face-centered cubic stacking (ABC sequence) as for the Ag substrate. This type of anisotropic effect has been studied in details to explain the growth of trigonal islands of Co/Pt(111) [21,22]. Indeed, the trigonal shape was explained by the anisotropic diffusion of adsorbed atoms along the step-edges (micro-facets) of the islands, which is more favorable for A-type micro-facets than the B-type. On the basis of this approach, one could deduce that an exchange between Zn and Ag atoms is more or less favorable depending on where the exchange occurs (on an A-type or B-type micro-facet) as explained in the 3D schemes of Fig. 6. This mechanism is only possible for a Zn substitution into the Ag plane.

We do not know which environment (A or B) is most favorable. The main point is that the two environments are very different from each other, and this could explain the preferential growth direction. On Fig. 6, we schematically present the expected shape of dendrites with the arbitrary assumption that the Zn substitution into the topmost Ag(111) layer is more favorable on the B-type micro-facets. One can imagine that the Zn ML growth that forms the MP is initiated by the formation of small nucleus of Zn atoms embedded in the step-edge and/or embedded in the topmost surface layer. Since the growth of Zn is assumed to be faster for B-type micro-facets, then the domains take a finger-like shape. Consequently, the segregation process of Zn (coming from the surface selvedge) results in the dendrites growth along the three directions. The related shape dynamics for the two types of step-edges are schematically plotted in Figs. 6d and 6e, respectively.

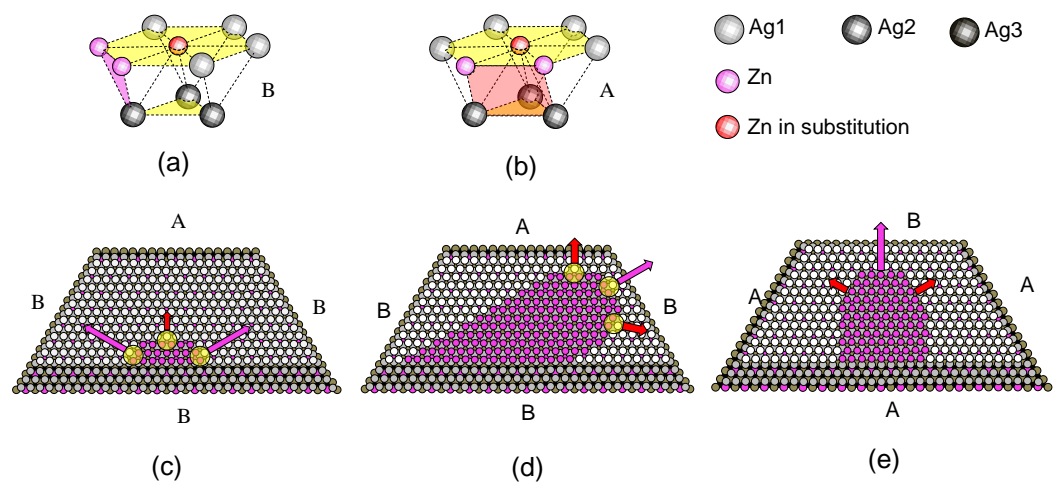

**Fig. 6.** Schematic presentation of anisotropic 2D dendritic growth in the case of the B-type micro-facet which is energetically more favorable than on the A-type. 3D schemes showing the structural difference between the Zn growth (a) on an A-type micro-facet and (b) on a B-type micro-facet; (c) Zn isotropic shape on a step-edge; (d) Zn anisotropic growth on a B-type step-edge; (e) Zn anisotropic growth on an A-type step-edge. In both cases, the growth is stopped on a perpendicular direction of a B-type micro-facet, due to the mismatch between the Ag and Zn (pseudomorphy strain relaxation).

For instance, at a low coverage of Ag atoms on the Pt substrate, at low temperature and lower Ag flux, the growth of two-dimensional Ag dendritic fractals is explained and simulated by classical Diffusion Limited Aggregation (DLA) mechanism [7]. The Ag dendritic fractals are formed on the upper surface of the substrate. These are main differences with respect to the Zn inserted dendrites.

For an equivalent low coverage of Co atoms on the Pt(111) substrate, surface dendrites are observed on the STM images [18,23]. The analysis at the atomic scale performed by Lundgren et al. [24] showed that, at first, the Co atoms are embedded into the Pt substrate, as in the case of Zn/Ag(111). However, the observed dendrites consists of Pt atoms onto the surface and consequently they are not the result of a DLA mechanism [25,26]. In addition, the Pt dendrites observed on the Co/Pt(111) are not inserted in contrast with the dendrites described here for the Zn/Ag(111) system. Thus, the results on dendrites growth we report here are original since no other similar growth mechanism was experimentally observed and no clear mechanism has been formulated up to now.

On the other hand, the Co monoatomic layer is formed on the Pt step-edge as in the Zn/Ag case. The main differences is that the Co monolayer is collinear with the Pt(111) substrate with extended (pseudomorphy) and relaxed (stacking fault) areas. However, in our case, the Zn monolayer is rotated by 1.5° with respect to Ag(111) with a large angular distortion (± 0.8°) [17]. The strain relaxation of the Zn ML does not form a periodic stacking fault, but a small rotation of the Zn ML (1.5°) with respect to Ag(111) in comparison with Co/Pt for which a quasi-pseudomorphy is observed.

From a practical perspective, the Co deposition on Pt(111) is studied for its interesting properties, particularly in the domain of magnetism. Despite the similarity between the Co/Pt(111) and the Zn/Ag(111) systems, the latter has drawn less interest due to its limited applications at present. However, the Zn/Ag(111) offers an alternative approach to gain more insight into the fabrication mechanism of self-assembled structures.

### 4.2. Transition from dendrites to Moiré pattern

At this point of the discussion, we exclusively discussed the evolution of the Moiré pattern on the dendritic side. The question to address is to define the role of the dendrites in the growth phenomenon of the dense Zn ML. It is not easy to clear up this question only by means of STM images, which display just frozen frames of the growth kinetics. In addition, we don't know where the step-edge is exactly located before Zn deposition, and consequently, what is the contribution of Zn dendrites to the formation and the evolution of the Zn MP. If one assumes that the Zn growth starts on the Ag step-edge in agreement with Fig. 4, therefore the orientation of the step-edge plays an important role on the formation and on the evolution of the dendrites.

The probability of the direct MP formation further to the advance of the MP-Ag front is insignificant, because the growth mechanism of the MP-Ag front is the same as the dendritic growth mechanism (*i.e.* the Zn substitution). In fact, in line with the growth mechanism proposed for dendritic growth, the Zn substitution is only possible on the B-type micro-facets by a perfect epitaxy. In this respect, one should consider the two extreme cases for different step-edge orientations.

When the step-edge is oriented at about 30° with respect to the dendritic growth direction, the MP-Ag interface cannot advance because it corresponds to the quasi A-type micro-facet (*e.g.* interface (1) in the Fig. 3a). A ball model of the dendrites growth and the dendrites-MP transition in this case is given in Fig. 7. The length of dendrites increases while the dendritic-width-induced strain is eliminated in the Ag inter-dendritic area. Thus, the inter-dendritic distance is regulated by the strain relaxation process. Besides secondary arm cross-back mechanism, the existence of two growth orientations of primary dendritic arms (with a 120° divergence) promotes the formation of Ag islands (I) by dendritic arms direct cross mechanism. This is a preliminary step before the formation of a Moiré pattern (Fig. 4). With the increase of Zn amount by substitution, the size of islands decreases and the dendrite width increases concomitantly. Due to its large surface area, the Zn is probably highly-strained and could be relaxed by increasing its local density, *i.e.* by changing its surface structure along with a small local angular shift. This is the probable reason for which the Moiré pattern behind the MP-Ag transition front is not perfectly regular and the angular dispersion of the Moiré pattern is very high, and exceeds 50% (1.5° ± 0.8°) [17].

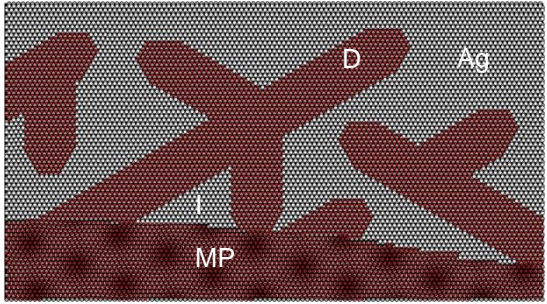

**Fig. 7.** Schematic presentation of 2D dendritic growth and D-MP transition on the quasi-A-type micro-facet showing the Ag island formation by secondary arm cross-back mechanism. Zn in the dendrite is in perfect epitaxy with Ag(111). The Zn monolayer in MP has an interatomic distance of 0.266 nm and is shifted of 1.5° relatively to Ag. MP: Moiré pattern, D: Dendrite, Ag: Silver terrace, I: Ag island.

When the step-edge is quasi-perpendicular to the primary dendritic arm orientation, the MP-Ag interface orientation is quasi-B-type and is favorable for the progress of MP-Ag interface. However, the direct advance, by Zn substitution, of the MP-Ag interface is insignificant with respect to dendritic growth due to the strain effect. The inter-dendritic distance stays constant along the dendrites growth direction since there is only one primary dendritic arm direction; the dendrites are thereby very parallel and regular. Consequently, the probability of Ag island formation by dendritic cross-back mechanism is insignificant. For this reason, the dendrites which are perpendicular to the MP-Ag interface are longer than those at 30°.

When the step-edge is in an intermediate orientation, for a relatively small domain, the evolution of dendrites leads to the formation of both A-type MP-Ag interface without dendrites and B-type MP-Ag interface with parallel regular dendrites. The two types of interfaces form an angle of 120° (Fig. 3). For relatively large domains, we can see a smooth variation of the configuration as shown in the typical example on Fig. 2.

The results obtained on the early-growth stages of Zn on Ag(111) can be discussed at the light of what is known on the nanoscale dendritic growth (nucleus segregation, crystal anisotropy, and growth anisotropy). The highly-anisotropic shape of Zn dendrites, their regular width and their confinement in the upper surface are the major advantages in the domain of the self-assembled nanotechnology.

## 5. Conclusions

Using AES and STM techniques, we have investigated the first growth stages of the Zn ML on Ag(111). We confirmed a dissolution process of the first deposited Zn atoms in the Ag surface selvedge. STM images recorded at different places of the sample revealed a growth mechanism of the Zn ML via the formation of embedded dendrites of Zn on the topmost Ag layer that has not been reported previously.

The explanation of the inserted dendritic growth mechanism is based on the STM observations and the segregation phenomena involving the competition between the size, the surface energy and the tendency to order. After the Zn atoms dissolution into the Ag substrate due to the size effect, the Zn is segregated in the surface on the upper border of the step edges. In contrast with classical layered growth, an anisotropic dendritic growth is established by Zn-Ag exchanges [15,27,28]. The non-center-symmetric growth comes from the anisotropy of the Ag(111) surface due to the ABC stacking configuration. Growth kinetics of dendrites has the threefold symmetry and do not conform to the symmetry center of crystal. This growth gives rise to a (1x1) superstructure which is strained only in one direction, limiting then the dendrite width to 2.5 nm. This confinement effect is due to the Zn anisotropic shape.

In the opposite direction of the dendritic growth, a Zn ML is formed by step-edge growth. In contrast with dendrites, the Zn ML conserves the symmetry of order 6 since it takes an isotropic shape. In this configuration, the strain relaxation occurs in two directions by Zn monolayer rotation with respect to the Ag(111) surface, which gives rise to a Moiré pattern.

**Acknowledgments**

We wish to thank Dr. G. Tréglia for fruitful discussions.